\documentclass[a4paper,fleqn,usenatbib]{mnras}

\usepackage[T1]{fontenc}
\usepackage{ae,aecompl}
\usepackage{rotating}

\usepackage{graphicx}	% Including figure files
\usepackage{amsmath}	% Advanced maths commands
\usepackage{amssymb}	% Extra maths symbols

\title[Plage regions and transits ]{The impact of faculae on the
  radius determination of exoplanets: The case of the M-star GJ\,1214.
  \footnote{Based on observations made with ESO Telescopes 
  at the La Silla Paranal Observatory under programme 099.C-0175(A)}}

\author[Eike W. Guenther]
  {E.W. Guenther, $^{1}$\thanks{E-mail: guenther@tls-tautenburg.de}
  \\
$^{1}$ Th\"uringer Landessternwarte Tautenburg, Sternwarte 5, 07778
  Tautenburg, Germany \\
}

\date{Accepted 12 January 2022. Received 6 January 2022; in original form 2 June  2021}

\pubyear{2021}

\begin{document}
\label{firstpage}
\pagerange{\pageref{firstpage}--\pageref{lastpage}}
\maketitle


\begin{abstract}
Precise measurements of exoplanets radii are of key importance for our
understanding of the origin and nature of these objects.  Measurement
of the planet radii using the transit method have reached a precision
that the effects of stellar surface features have to be taken into
account.  While the effects from spots have already been studied in
detail, our knowledge of the effects caused by faculae is still
limited. This is particularly the case for M-stars. Faculae can pose a
problem if they are inhomogeneously distributed on the stellar
surface. Using the eclipse mapping method, we study the distribution
of the faculae on the surface of GJ\,1214 using the Ca\,{\tiny
  II}\,H\&K lines as tracers. In order to assess the homogeneity  of the
distribution in a quantitative way, we introduce the inhomogeneity
factor IHF. IHF is 0\% if the distribution is homogeneous, positive, if
the plage regions are preferentially located along the path of the planet,
and negative, if they are preferentially located outside the path 
of the planet.  For GJ\,1214, we derive a
rather small value of $\rm IHF=7.7_{-7.7}^{+12.0}\%$. We discuss the
relevance of this result in the context of the PLATO and ARIEL
missions.
\end{abstract}
  


%%%%%%%%%%%%%%%%%%%%%%%%%%%%%%%%%%%%%%%%%%%%%%%%%%

%%%%%%%%%%%%%%%%% BODY OF PAPER %%%%%%%%%%%%%%%%%%

\section{Introduction}
\label{sectI}

% \bf --> \revision
 
In recent years the focus of exoplanet research has shifted from the
mere detection  to their detailed characterization.
The detailed characterization requires accurate determinations of
their masses, radii and studies of their atmospheres.  As up to now,
transiting exoplanets are the best objects to obtain such information.

The combination of precise radius- and mass-determinations
of exoplanets constrains the composition of the planets
\citep{neil2020, kuchner2003, leger2004, zeng2019, adams2008}.
\citet{fridlund2020} estimates that an accuracy of better than 15\%
for the masses and better than 5\% is required to find out what the
nature of planets are. The requirement for the
radius determinations thus are more stringent than for the mass
determinations. Such considerations are the background of the 
upcoming PLATO mission \citep{rauer2014}. 

Key information is also provided by atmospheric studies such as those
that will be obtained with the atmospheric characterization mission
ARIEL \citep{tinetti2016,puig2018}. PLATO and ARIEL will both
primarily use the transit method to measure planet radii to very
highes accuracies.  In the case of ARIEL, these are obtained at many
different wavelengths in the optical and infrared regime.

The accuracy with which the radii can be determined not only depends
on the instrument, it also depends on how well the effects from stellar
surface features can be removed. It
is thus essential to develop these methods.

Stellar spots that are being occulted during transit
reduce the transit depth, the planet appears to be smaller than
it really is.  Spots may form bands on the stellar
surface and if the planet transit along such a band, it may become
very difficult to account for the effects of spots. Sophisticated
methods are then need to determine the true size of the planet
\citep{morris2018}.

An example for a star with two active bands is HAT-P-11
  \citep{morris2017}. Because the planet is misaligned, it is crossing
  the bands during transit \citep{sanchis-ojeda2011}. If the planet
  were aligned, and if it would transit at the latitude of the bands,
  it would be the situation as envisaged by \cite{morris2018}. 

Stars can also have spots that are not occulted dring transit.  In
this case the planet appears to be larger than it really is. This
situation would, for example, be the case if the star has polar spots
and the planet transits close to the equator. This situation is
particularly relevant for M-stars, because they have polar spots.
Using white-light observation of long-duration flares of fully
convective M-stars \citet{ilin2021} showed that all flares must have
occurred at latitudes between 55 and 81 degrees. {Zeeman Doppler
  imaging shows that M-stars have a complex field topology including
  polar spots \citep{donati2008, morin2008, morin2010, barnes2015}.

Stars not only have spots but also faculae, or to be more precise,
facula points.  Facula points pose a problem if they are clumped on
the stellar surface.  Studies of faculae on the Sun show that they
consist of many small magnetic elements.  Faculae are bright, because
of the ``hot wall'' effect \citep{kobel2009}.  Near the disk center,
they appear as {\em bright points} or {\em filigree} \citep{dunn1973,
  mehltretter1974}.  When observed near the limb, they are called
faculae, or more precisely, {\em facular grains}
\citep{muller1975}. Bright points, filigree and facular grains are
thus different manifestations of the same phenomena \citep{kobel2009}.
Regions that appear bright in the cores of the Ca\,{\tiny II} lines
are called {\em plage regions}.  The plage regions are also related to
the small magnetic elements. Numerical simulations show that the
emission cores of the Ca\,{\tiny II} lines are the result of the
acoustic wave heating and magnetic wave heating from magnetic elements
\citep{fawzy2002a,fawzy2002b,fawzy2002c}.  The emission cores of the
Ca\,{\tiny II} lines are thus the chromospheric signatures of the
magnetic elements which we called bright points, filigree, or facular
grains when observed in the photosphere.

Faculae are observed at the solar limb, because we observe a slightly
higher layer at the limb than at the disk center \citep{kobel2009}. At
the disk center we either observe them as facular grains in the
photosphere, or as plages in in the chromosphere using Ca\,{\tiny
    II} lines.

Because faculae and plage regions are related, and since we use the
Ca\,{\tiny II} lines in our study, we will call these regions plages.

Plage regions can also be observed on other stars by using the
Ca\,{\tiny II} lines.  The presence of Ca\,{\tiny II}\,H\&K emission
shows that even M-stars have plage regions. Furthermore, bright
regions on the stellar surface have been detected on the late M-star
TRAPIST-1 \citep{morris2018a}. Plage regions thus seem to exist on all
late-type stars including M-stars. However, our know\-ledge 
  about plage regions on other types of stars is still limited,
particularly for M-stars.  Studying the effects of plage regions on
M-stars is important, because many known low-mass planets orbit 
such stars.  In the following we will also call the Ca\,{\tiny
    II} emitting regions on M-stars plage regions. Because M-stars
can have polar spots, they can also have polar plage regions.

In this article we will show how the transit mapping method using the
Ca\,{\tiny II}\,H\&K lines can be used to determine how homogeneously
the facula points are distributed on the stellar surface.  We select the M-star 
GJ\,1214 as a benchmark object, because it is one of the best studied M-stars
hosting a planet.

In Section~\ref{sectII} we summarise what is known about GJ\,1214 and
its planet. In Section~\ref{sectIII} we explain that the Ca\,{\tiny
  II}\,H\&K lines can be used to find out how homogeneous the
distribution of plage regions is. In that section we also introduce
the inhomogeneity factor (IHF).  The observations are presented in
Section~\ref{sectIV}.  In in Section~\ref{sectV} we determine the
filling factor of GJ\,1214.  IHF is determined in
Section~\ref{sectVI}.  The impact of faculae on transit observations
obtained with PLATO and ARIEL are finally discussed in
Section~\ref{sectVII}.

\section{The target: GJ\,1214}
\label{sectII}

For our study we selected GJ\,1214 which is a bright, nearby M4.5-star
that harbors a transiting planet \citep{charbonneau2009}. The mass,
radius, temperature and luminosity of this star are: $\rm
M_*=0.176\pm0.0087,M_{\odot}$, $\rm R_*=0.213\pm0.011\,R_{\odot}$,
$\rm T_{eff}=3252\pm201\,K$, and $\rm L=(4.05\pm0.19)\times
10^{-3}\,L_{\odot}$, respectively.  The mass, radius and bulk density
of the planet are: $\rm M_p=6.26\pm0.91\,M_{\oplus}$, $\rm
R_p=2.80\pm0.24\,R_{\oplus}$, and $\rm \rho = 1.56\pm0.61\,gcm^{-3}$,
respectively \citep{anglada2013}.

GJ\,1214b is a benchmark object, because the mass and radius of this
planet can be obtained with high accuracy.  \citet{puig2018} also used
GJ\,1214b as an example for an M-star planet that is going to be
observed with ARIEL.  

A large number of transit observations using satellites, airplanes,
and ground-based telescopes of have already been obtained
\citep{bean2010, bean2011,berta2011,croll2011,desert2011,
  murgas2012,fraine2013,gillon2014,narita2013,kreidberg2014,wilson2014,
  nascimbeni2015,angerhausen2017,rackham2017}.  All these observations
show a flat, featureless spectrum. Not even the Helium line has
been detected in high resolution observations \citep{kasper2020}. The
planet could either have a rocky core and a hazy atmosphere, or it
could also be a multi-component water world \citep{roger2010,
  zeng2019}. GJ\,1214b thus belong to the class of exoplanets were the
currently available data does not allow to resolve its nature. It is
thus a key target of future observations.

One possibility to solve this mystery would be the detection of the
Rayleigh scattering its atmosphere.  The detection of Rayleigh
scattering could be possible even if the atmosphere is hazy.  A
tentative detection was presented by \citet{demooij2012} but this
detection was not confirmed with observations of higher sensitivity
\citep{nascimbeni2015}.  However, one problem of the attempts to
detect Rayleigh scattering in the atmosphere of the planet is that
faculae on the stellar surface can mimic Rayleigh scattering
\citep{oshagh2014}. Faculae can thus affect the diameter measurements
of the planet as well as studies of its atmosphere.

The spots of GJ\,1214 have already been intensively studied using
broad band photometry \citep{narita2013,rackham2017}. Possibly, the
most comprehensive study of this kind was carried out by
\citet{mallonn2018,mallonn2019} who monitored the star since
2014. \citet{mallonn2018} determined an average spot filling factor of
$0.25\pm0.09$\%, a temperature contrast of the spots of $\sim$ 370 K
and a rotation period of $\rm P_{star} =125\pm5$ days.
\cite{schlaufman2010} identify exoplanet systems that are likely to be
misaligned using statistical methods. Since GJ\,1214b is not
  amongst these systems, it is unlikely to be misaligned. According
to \citet{berta2011} the inclination of GJ1214 is
$88.80^{+0.25}_{-0.20}$ degrees.  We thus view this star nearly
equator on.

The effects of spots have already been taken in to account in the
analysis of the transit observations.  The aim of this work is to find
out what the effect of faculae are.  This is important for future
observations, including attempts to detect the Rayleigh scattering.
Since GJ1214 photometrically variable, not all spots are
at the poles. Since spots and plage regions are usually related, 
it is not likely that all plage regions are at the poles.
It is thus possible that the crossing of plage
regions could also affect the transit light-curve.

\section{The inhomogeneity factor of plage regions}
\label{sectIII}

\subsection{The definition of the inhomogeneity factor (IHF) }

Using the transit mapping method, \citet{wolter2009} studied the
locations of stellar spots along the path of the planet for CoRoT-2b.
The same method can also be used to map out the location of plage
regions along the path of the planet using the Ca\,{\tiny II}
lines. In this way we can find out whether the plage regions are
clumped, or homogeneously distributed on the stellar surface. A clump
of plages would appear as a hump in the transit light-curve in the
Ca\,{\tiny II} lines, just as \citet{wolter2009} observed a hump in
the continuum caused by spots.

By comparing the strength of the Ca\,{\tiny II} lines in- and out-of
transit, we can find out whether the filling factor is the same along
the path of the planet as on the rest of star. There are three
different cases:

\begin{itemize}
\item A.) The plage regions are homogeneously distributed on the
  stellar surface: In this case, the continuum emission and the
  emission in the line cores of the Ca\,{\tiny II}\,H\&K would both
  decrease by the same amount during transit.  After normalising the
  spectra to the continuum, the relative line fluxes {\em would be the
    same}.
\item B) The plage regions are inhomogeneously distributed on the
  stellar surface, and preferentially located along the path of the
  planet: In this case, the decrease of Ca\,{\tiny II}\,H\&K fluxes is
  larger than the decrease of the continuum. This means that the
  normalised Ca\,{\tiny II}\,H\&K fluxes would {\em decrease} during
  transit.
\item C) The plage regions are inhomogeneously distributed on the
  stellar surface but located outside the path of the planet. In this
  case the Ca\,{\tiny II}\,H\&K fluxes would not change but the
  continuum flux would decrease.  This normalised Ca\,{\tiny II}\,H\&K
  fluxes would thus {\em increase} during transit.
  Case C is more likely than case B, since the area occulted 
  by the planet is usually small compared to the total area of the
  plage regions.
\end{itemize}

If A is the case, the diameter measurements of the planet are not be
affected by plage regions. This would, for example be the case if the
filling factor is either 100\%, or 0\%. A star without without active
regions could still have some emission in the Ca\,{\tiny II}\,H\&K
lines, because of the acoustic heating.  As we will
show in the next section, the filling factor of GJ\,1214 is neither
100\%, nor 0\%. If B is the case, the transit observations are
affected.

In order to visualise the cases B and C, one may think of a planet
that transits along equator of a star that is viewed equator-on.  In case
B the plage regions would be close to the equator, and in case C close
to the poles. 

For quantifying the impact of plage regions, we have to distinguish
two possibilities: $\rm A_{plage}> A_{planet}$ and $\rm A_{plage}<
A_{planet}$, with $\rm A_{plage}$ the area of the plage regions on the
stellar surface, and $\rm A_{planet}$ the area occulted by the planet
during transit.

To explain why it matters if the plage regions are larger or smaller
than the planet let us assume case B and $\rm A_{plage}=10 \cdot
A_{planet}$.  The maximum decrease of the Ca\,{\tiny
  II}\,H\&K lines would then be 10\%.  However, if $\rm A_{plage}\leq
A_{planet}$, the Ca\,{\tiny  II}\,H\&K line would completely disappear during transit,
if there is one plage region which is located along the path of the planet.
Thus, in one case a 10\% decrease would indicate the
maximum inhomogeneity, in the other, a 100\% decrease. We 
have to distinguish two cases:

\begin{equation}
{\rm For\,A_{plage}\geq A_{planet}:
IHF = \frac{\left(1- \frac{CaII_{IT}}{CaII_{OT}} \right)}{\left(
    \frac{A_{planet}}{A_{plage}} \right)}}
\label{eq01}
\end{equation}

\begin{equation}
{\rm For\, A_{plage}\leq A_{planet}: 
IHF = 1- \frac{CaII_{IT}}{CaII_{OT}}  }
\label{eq02}
\end{equation}

where $\rm {CaII_{IT}}$ is the flux of the Ca\,{\tiny II}\,H\&K lines
measured during transit, and $\rm {CaII_{OT}}$ the flux out-of
transit.  $\rm A_{plage}$ and $\rm A_{planet}$ are the surface areas
on the star covered by plage regions and the planet during transit.
Note that $\rm CaII_{IT}\leq CaII_{OT}$ is always fulfilled for case
B. For $\rm A_{planet}=A_{plage}$ both equations are the
same.

IHF=100\% means the distribution is as inhomogeneous as possible, for
example if there is just one plage region on the star.  IHF=0\% means
that the distribution is homogenous (case A), or at least does not
affect the transit observations.

Let us assume for case C a similar situation as above, $\rm
A_{plage}=10 \cdot A_{planet}$ and a planet that covers 1\% of the
surface of the star. During transit, the flux of the Ca\,{\tiny
  II}\,H\&K lines would not change but the flux of the other regions
would decrease by 1.1\%. After normalising the spectra, the Ca\,{\tiny
  II}\,H\&K lines would increase, because the continuum decreases
during transit.  Like for spots, the effect of non-occulted plage
regions is smaller than that of occulted ones. 
IHF is negative in case C. 

As will be demonstrated in Section \ref{sectVII} for the case of
GJ\,1214b, the IHF-value can be used to determine the correction
factor that has to be applied in order to correct the diameter
measurements of the planet for a star with plage regions.

\subsection{Comparing the IHF-method with other approaches and special cases
of transits}

An alternative method has been developed by \cite{morris2018}. In this
approach, the ratio of $\rm p_0=R_p/R*$ to $\rm p_1=\sqrt
\delta$,  is derived ($\rm \delta$ is the transit depth.). The
basic idea is to determine $\rm p_1$ from the light-curve and $\rm
p_0$ using other methods.  For determining $\rm p_1$, the impact
parameter is estimated using the dependence of the transit duration
upon the stellar density and the eccentricity of the planet's orbit,
as well as other quantities such as the planet's orbital period.  This
in turn requires to determine the stellar density independently from
the light-curve modeling. As \cite{morris2018} pointed out,
astroseismology could provide this information. The method thus can be
used for solar-like stars were the density of the star has been
determined using astroseismology. However, the amplitudes of stellar
oscillations on M-stars are tiny \citep{kjeldsen1995}.  It may thus be
quite difficult to obtain the stellar density from astroseismology for
M-stars.

The IHF-quantity has the advantage that it is directly derived from
the observations, without making any additional assumptions, or the
need of additional information coming from other sources.  It also
works for M-stars, and for stars with activity bands.  Since the
Ca\,{\tiny II} emission would decrease during transit, it would be
obvious that the planet has occulted active bands during transit.
  
The IHF-quantity can also be used to derive $p_0/p_1$.  All we need
is  the temperature difference between the plage regions for that (See
Section~\ref{sectVII}).
  
If the planet orbits over the poles and occults active regions located
at the north and the south pole, it would also be case B.
  
The IHF-quantity has also the advantage that it is proportional to the
change of the depth of the transit due to plage regions.  For example,
if the depth changes by 100 ppm for IHF=10\% then it would change by
200 ppm for IHF=20\%.  The IHF quantity thus is a measure how much the
transit depth is changed because of plage regions.

\section{Observations and data reduction}
\label{sectIV}

We observed GJ\,1214 continuously from July 29, 2017 UTC 23:34 until
July 30, 2017 UTC 03:25 with UVES at the VLT (ESO program
099.C-0175(A)). The midpoint of the transit was at $\rm T_c=
BJD\,2457964.552073\pm0.00032$ \citep{kasper2020}.  The middle of the
transit was on July 30, 2017 at UTC 01:15. The transit lasted from
00:49 to 01:41 The ingress and egress takes only 6 min. During the 3
hours and 51 minutes of observations, we obtained 18 spectra with
exposure times of 12 min.  We obtained five spectra before the
transit, five during the transit, and eight spectra after it. The
observations before the transit were obtained at airmass 1.31 to 1.18,
during transit at airmass 1.18 to 1.15, and after it at airmass 1.15
to 1.29.  The spectra cover the wavelength range from 325.9 to 449.3
nm in the blue and 472.6 nm to 683.5 nm in the red arm of the
spectrograph.  The resolution of the spectra is $\lambda/\Delta
\lambda = 52000$ with the 0.8 arcsec slit used.  The standard ESO
reduction pipeline was used. We also reduced the spectra also with
IRAF to make sure that the results do not depend on the data-reduction
process.

\section{The plage filling factor}
\label{sectV}

For deriving the inhomogeneity factor (IHF), we have to find out
whether $\rm A_{plage}> A_{planet}$ or $\rm A_{plage}\leq A_{planet}$
(equation~\ref{eq01} and \ref{eq02}).  For determining the filling
factor, we use the flux of the Ca\,{\tiny II}\,H\&K.  Solar and
stellar observations show that the unsigned magnetic field strengths
and the flux of the Ca\,{\tiny II}\,H,K lines are related
\citep{schrijver1989,loukitcheva2009,morgenthaler2012,chatzistergos2019}.
This relation does not depend on the magnetic cycle and it can be
applied to all G, K, and M-stars. The relation is linear for large
magnetic field strength. For weak fields, it becomes non-linear.
Solar observations of high spatial resolution show that there are two
sources of the Ca\,{\tiny II} flux.  A basal flux of nonmagnetic
origin, and a component that is related to the magnetic field
\citep{loukitcheva2009}.  Theoretical studies show that the basal flux
is due to the acoustic heating of the chromosphere \citep{fawzy2002a,
  fawzy2002b, fawzy2002c}.  Comparing theory with observations,
\cite{fawzy2002b} find that the relative contribution of the acoustic
heating decreases with decreasing mass of the star.  This means, the
basal flux is less important or M-stars than for solar-like stars.
  
Because our aim is to find out what the maximum effect of plage
regions could be, it is assumed that the flux of the Ca\,{\tiny
  II}\,H\&K is only due to active regions. If some of it is of
nonmagnetic origin, the effects due to plage regions would be smaller.
Strictly speaking, the filling factor (ff) thus is an upper limit,
because some of the flux of the Ca\,{\tiny II}\,H,K lines could be due
to acoustic heating.

The (upper limit of the) filling factor (ff) is determined from the
ratio of the flux of the Ca\,{\tiny II}\,H\&K of GJ\,1214 to the
reference star (ref) with a known filling factor:

\begin{equation}
{\rm ff_{GJ1214} \leq \frac{F_{CaII,GJ1214}} {F_{CaII,ref}}*ff_{ref}}
\label{eq03}
\end{equation}

A reference star is a star of the same spectral type as the target
were the filling factor and the magnetic field strength has been
measured.

We use AD\,Leo (=GJ 388) as a reference, because there are several
measurements of the filling factor and the magnetic field strength.
AD\,Leo is an M4.5 star with a rotation period of 2.23 days and $\rm
T_{eff}=3414\pm100\,K$ \citep{DiMaio2020}.  \citet{saar1985} derived
$\rm ff=73\pm6\%$ and $\rm B=3800\pm260\,G$.  \citet{cranmer2011}
found $\rm ff=60\%$ and $\rm B=4000\,G$. \citet{shulyak2017}
determined a magnetic field strength of $\rm
ffB=3100^{+100}_{-200}\,G$.

The total flux of the Ca\,{\tiny II}\,H lines of GJ\,1214 is $\rm
(13.0\pm0.5)\times 10^{24}\,erg\,s^{-1}$ and that of the Ca\,{\tiny
  II}\,K line is $\rm (8.0\pm0.5)\times 10^{24}\,erg\,s^{-1}$.  Using
these values, the Ca\,{\tiny II}\,H,K fluxes of AD\,Leo, and taking
also the slight differences between the two stars in to account, we
obtain $\rm ff\leq 6.8\pm1.0$\% for GJ\,1214.

\cite{fawzy2002b} calculated the flux of the Ca\,{\tiny II}\,H,K for
stars due to acoustic and magnetic wave heating for stars with
spectral types in the range between F5 and M0. GJ\,1214 is outside
this range but if we extrapolate this relation, we end up with filling
factors in the range between 5 to 10\%. The value of 7\% thus is
reasonable.
  
This means, we have to use equation~\ref{eq01} for calculating IHF,
because the planet occults only $\rm 1.39\pm0.25\%$ of the stellar
surface.  We find $\rm A_{planet}/A_{plage}=0.204\pm0.047$.  Since the
spot filling factor is only $0.25\pm0.09$\% \citep{mallonn2018}), most
of the magnetic flux is in the plage regions.

\section{The homogeneity of the plage regions}
\label{sectVI}

The pseudo-Equivalent Width (pEW) of GJ\,1214 obtained are given in
Table~\ref{tab:EWa}. We prefer to call these pseudo-equivalent width,
rather than equivalent width, because we measure only the equivalent
width the emission core. Fig.\,\ref{HKEW} shows the normalized values
obtained during the night. We normalized the values for this figure,
so that both lines can be shown in the same figure.

\begin{table*}
\caption{Pseudo equivalent width (pEW)}
\begin{tabular}{c c c}
\hline
\noalign{\smallskip}
HJD-2457964 & pEW Ca\,II-H & pEW Ca\,II-K \\
                       & [\AA ]             & [\AA ]  \\
\hline
0.48976 & $-1.19\pm0.10$ & $-2.65\pm0.17$ \\
0.49863 & $-1.06\pm0.07$ & $-3.05\pm0.17$ \\ 
0.50755 & $-1.03\pm0.05$ & $-2.73\pm0.19$ \\
0.51647 & $-1.06\pm0.11$ & $-2.60\pm0.29$ \\ 
0.52539 & $-1.27\pm0.11$ & $-3.26\pm0.30$ \\ 
0.53431 & $-1.04\pm0.11$ & $-2.64\pm0.27$ \\ 
0.54326 & $-1.06\pm0.05$ & $-2.73\pm0.10$ \\ 
0.55215 & $-1.02\pm0.05$ & $-2.89\pm0.10$ \\ 
0.56107 & $-1.08\pm0.06$ & $-2.98\pm0.12$ \\
0.57002 & $-0.93\pm0.09$ & $-2.70\pm0.12$ \\
0.57899 & $-1.14\pm0.13$ & $-2.80\pm0.10$ \\ 
0.58785 & $-1.24\pm0.05$ & $-2.86\pm0.20$ \\
0.59677 & $-1.25\pm0.05$ & $-3.25\pm0.26$ \\
0.60570 & $-1.22\pm0.09$ & $-2.63\pm0.28$ \\
0.61462 & $-1.06\pm0.10$ & $-2.72\pm0.31$ \\
0.62354 & $-1.30\pm0.10$ & $-3.32\pm0.29$ \\
0.63247 & $-1.12\pm0.07$ & $-3.33\pm0.20$ \\
0.64140 & $-1.24\pm0.10$ & $-3.30\pm0.17$ \\
\end{tabular}
\label{tab:EWa}
\end{table*}

\begin{table*}
\caption{Average pseudo equivalent width (pEW)}
\begin{tabular}{l c c c c c c}
\hline
\noalign{\smallskip}
phase & Ca\,II-H$^{1}$ & Ca\,II-H$^{1}$ & Ca\,II-K$^{1}$ & Ca\,II-K$^{1}$ & Ca\,II-HK$^{1}$ & Ca\,II-HK$^{1}$ \\
      &                & trend sub.$^2$ &                & trend sub.$^2$ &                 & trend sub.$^2$ \\
      & [\AA ] & [\AA ] & [\AA ] & [\AA ] & [\AA ] & [\AA ] \\
\hline
$\rm pEW_{OT, before}$  & $-1.122\pm0.042$ & $-1.173\pm0.041$ & $-2.859\pm0.115$ & $-3.011\pm0.109$ & $-3.982\pm0.141$ & $-4.140\pm0.109$ \\
$\rm pEW_{IT}$         & $-1.027\pm0.022$  & $-1.038\pm0.025$ & $-2.792\pm0.056$ & $-2.827\pm0.052$ & $-3.819\pm0.069$ & $-3.956\pm0.052$  \\
$\rm pEW_{OT, after}$    & $-1.181\pm0.033$  & $-1.154\pm0.035$ & $-2.851\pm0.096$ & $-2.770\pm0.101$ & $-4.032\pm0.115$ & $-3.899\pm0.101$ \\
$\rm pEW_{OT}$      & $-1.152\pm0.028$  & $-1.163\pm0.027$ & $-2.855\pm0.075$ & $-2.890\pm0.084$ & $-4.007\pm0.092$ & $-4.019\pm0.084$ \\
\hline 
\end{tabular}
\label{tab:EW}
\\
$^1$ Five measurements before, during and after the transit used. \\
$^2$ trend subtracted. \\
\end{table*}

\begin{figure}
\centering
 \includegraphics[height=0.27\textheight,angle=0.0]{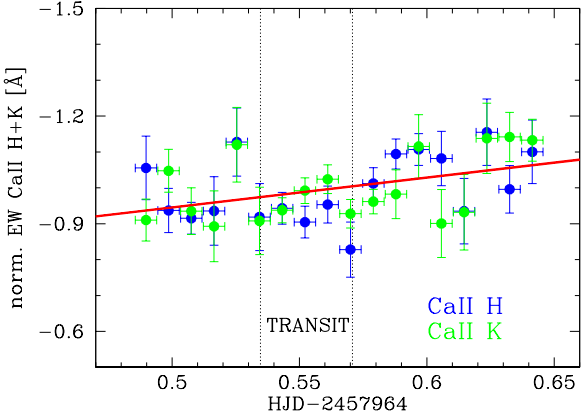}
\caption{Normalized equivalent widths of the Ca\,II-H (blue) and
  Ca\,II-K (green) lines of GJ\,1214. The red line shows the trend
  subtracted.}
  \label{HKEW}
 \end{figure} 

\begin{figure}
\centering
 \includegraphics[height=0.27\textheight,angle=0.0]{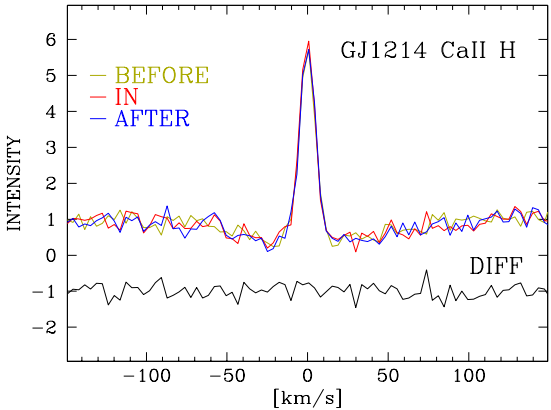}

\caption{Spectrum of the Ca\,II-H line of GJ\,1214, before (dark
  yellow), during (red) and after (blue) the transit.  Below is the
  difference between the spectra taken in- and out-of transit. There
  is no significant difference.}
\label{CaIIH}
 \end{figure} 

\begin{figure}
\centering
 \includegraphics[height=0.27\textheight,angle=0.0]{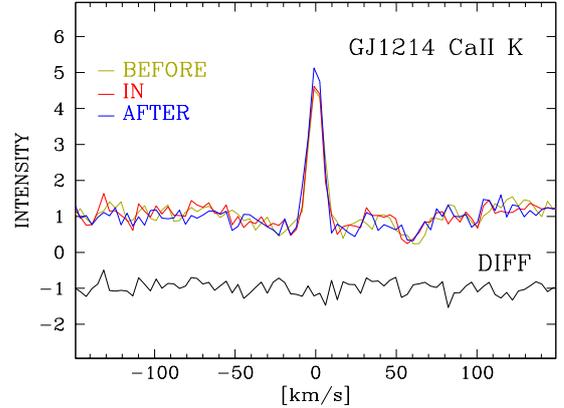}
\caption{Same as Fig.\,\ref{CaIIH} but for the Ca\,II-K line.}
  \label{CaIIK}
 \end{figure} 

The spectra taken before (dark yellow), during (red) and after (blue)
the transit are shown in Fig.\,\ref{CaIIH}, and Fig.\,\ref{CaIIK}. The
lines at the bottom of the figure are the difference between the
spectra taken in- and out-of transit. One may think that the
Ca\,{\tiny II}\,K line is higher after the transit then during transit
(Fig.\,\ref{CaIIK}). However, this difference is not significant.

Because both lines show exactly the same behavior, we sum up the pEWs
of both lines to increase the accuracy. We derive: $\rm
pEW=-3.982\pm0.141$7\,\AA\, before the transit, $\rm
pEW=-3.819\pm0.069$\,\AA\, during the transit, and $\rm
pEW=-4.032\pm0.115$\,\AA\, after it.

Both lines show a trend Fig.\,\ref{HKEW}. This trend is just the
signature of a small increase of the activity level of the star during
the night. After removing this trend, we obtain $\rm
pEW=-3.956\pm0.052$\,\AA \,during transit, and $\rm
pEW=-4.019\pm0.084$ out-of transit. The difference thus is
$0.063\pm0.099$\,\AA . Thus, there is no significant change of the
Ca\,{\tiny II}\,H,K lines during transit. GJ\,1214 thus corresponds to
case A in Section~\ref{sectIII}. All values derived are listed in
Table~\ref{tab:EW}.

Using the values with the trend subtracted, we obtain $\rm
IHF=7.7_{-7.7}^{+12.0}\%$.  The errors are relatively large, because
the planet occults only a small fraction of the plage regions.
Nevertheless, we conclude that there is no evidence that the plage
regions are clumped.

Our result thus is consistent with the results obtained by
\cite{gillon2014} that find no evidence that the transit chord of
GJ\,1214b is different in brightness to the rest of the stellar
surface.

\section{Discussion and conclusions}
\label{sectVII}

\citet{oshagh2014} modeled the effects of faculae that are occulted
during transit. In their model, they assumed an M-dwarfs of 3000 K and
planets with $\rm R_p/R_{star}=0.05$ to 0.15.  The authors furthermore
assumed that the faculae are 100 K hotter than the normal photosphere
and cover between 0.25 and 6.25\% of the surfaces of the star. The
values obtained for GJ\,1214 are thus within the range of this
model. However, \citet{oshagh2014} could not know, if the faculae are
distributed homogeneous or inhomogeneous on the stellar surface.

Using IHF, we can now calculate how large the affect for PLATO
observation would be, in other words, we can calculate $p_0/p_1$.
According to the numerical simulations by \citep{beeck2015}, the
average temperature difference between a region with a magnetic field
strength of 500 G and non-magnetic region is only 20 K for early-type
M-stars.

Let us assume that that the decrease of the strength of the
CaIIHK-lines is not an upper limit but a real decrease of 6.3\% during
transit. In this case the brightness of the star would change by 41
ppm.  Thus, faculae do not pose a big problem for the PLATO-, as well
as ARIEL-observations of GJ\,1214b.  However, the effects of faculae
could be much larger for other stars. It is thus useful to
observe the transits in Ca\,{\tiny II}\,H,K of all stars that ARIEL
will observe and also for the planet host stars that PLATO will
detect. Since the design goal of PLATO is a photometric accuracy of 34
ppm \citep{rauer2014}, plage regions may affect the diameter
measurements in some objects. The effects of faculae should thus be
included into simulations, like those presented by \citet{samadi2019}.

The method can also be generalized using other lines that trace only
active regions. For solar-like stars the CO-lines in the infrared,
which originate only from spots, would be an option.

In interesting aspect pointed out by \cite{isik2020} is that, at least
on solar-like stars, there are not only active latitudes but also
active longitudes. Such a clustering of active regions leads to an
enhancement of the photometric variability compared to random
distribution of active regions.

The detailed analysis of the impact of active regions remains to be an
issue for precise measurements of the diameters of planets. For the
PLATO mission, spectro-polarimetric observations are planed for the
most interesting targets (PLATO-WP\,145200). However, such
observations are very demanding and thus can only be carried out for
very few objects. Observations of the Ca\,{\tiny II}\,H,K represent a
cheap alternative and can be obtained for many stars (PLATO-WP\,11 and
WP\,14).

\section*{Acknowledgements}

The UVES observations were obtained in ESO programme 099.C-0175(A). We
are very thankful to the ESO-staff for carrying out the observations
in service mode, and for providing the community with all the
necessary tools for reducing and analyzing the data.  This work was
generously supported by the Deutsche Forschungsgemeinschaft (DFG) in
the framework of the priority programme ``Exploring the Diversity of
Extrasolar Planets'' (SPP 1992) in program GU 464/22, and by the
Th\"uringer Ministerium f\"ur Wirtschaft, Wissenschaft und Digitale
Gesellschaft.  This research has made use of the SIMBAD database,
operated at CDS, Strasbourg, France.

\section*{Data Availability Statement}

The data underlying this article are available in the ESO Science
Archive Facility http://archive.eso.org/cms.html.

%%%%%%%%%%%%%%%%%%%%%%%%%%%%%%%%%%%%%%%%%%%%%%%%%%



%\bibliographystyle{mnras}


% \bibitem[\protect\citeauthoryear{Author}{2012}]{Author2012}
% Author A.~N., 2013, Journal of Improbable Astronomy, 1, 1

% \bibitem[\protect\citeauthoryear{Others}{2013}]{Others2013}
% Others S., 2012, Journal of Interesting Stuff, 17, 198

% xxxxxxxxxxxxxxxxxxxxxxxxxxxxxxxxx

%%%%%%%%%%%%%%%%%%%%%%%%%%%%%%%%%%%%%%%%%%%%%%%%%%

%%%%%%%%%%%%%%%%% APPENDICES %%%%%%%%%%%%%%%%%%%%%

% \appendix


%%%%%%%%%%%%%%%%%%%%%%%%%%%%%%%%%%%%%%%%%%%%%%%%%%

% Don't change these lines
\bsp	% typesetting comment
\label{lastpage}
\end{document}